%Paper: hep-th/9305098
%From: "Chris M. Hull" <CMH@sbitp.itp.ucsb.edu>
%Date: Thu, 20 May 1993 20:37 PST

%%  %%%%%%%This requires the PHYZZX.TEX macropackage

\def\QQ{{Q}}
\def\f{{\phi}}
\def\g{{\gamma}}
\def\b{{\beta}}
\def\pa{{\partial}}
\def\ha{{1 \over 2}}
\def\pf{{\partial \phi}}
\def\dd{{\Delta}}
\def\si{{1 \over 16}}

 \tolerance=10000
  \input phyzzx.tex

 \nopubblock
%\Pubnum = {QMW-93-14}
%\date = {May 1993}
%\pubtype={}
{\begingroup \tabskip=\hsize minus \hsize
   \baselineskip=1.5\ht\strutbox \topspace-2\baselineskip
\halign to\hsize{\strut #\hfil\tabskip=0pt\crcr
{NSF-ITP-93-65}\cr
{QMW-93-14}\cr   {hep-th/9305098} \cr
   {May 1993}\cr }\endgroup}

\titlepage
\title {{\bf NEW REALISATIONS OF MINIMAL MODELS AND THE
STRUCTURE
OF W-STRINGS}}

\author {C. M. Hull}
\
\address {Physics Department,
Queen Mary and Westfield College,
\break
Mile End Road, London E1 4NS, United Kingdom.}

\andaddress{Institute for Theoretical Physics,University of California,
Santa Barbara, CA 93106, USA.}

\abstract
{ The quantization of a free boson whose momentum satisfies a cubic
constraint leads to a  $c=\ha$ conformal field theory  with a BRST
symmetry.  The theory also has a $W_\infty $ symmetry in which all the
generators except the stress-tensor are BRST-exact and so topological.
The BRST cohomology  includes states of conformal dimensions
$0,\si,\ha$, together with \lq copies' of these states  obtained by
acting with picture-changing  and screening operators. The 3-point and
4-point correlation functions agree with those of the Ising model,
suggesting that the theory is equivalent to the
critical Ising model. At tree level, the $W_3$ string can be viewed as
an ordinary $c=26$ string whose conformal matter sector includes this
realisation of the Ising model.  The two-boson $W_3$ string is
equivalent to the Ising model coupled to two-dimensional quantum
gravity. Similar results apply for other W-strings and minimal models.}

 \endpage

\pagenumber=1

 \REF\zam{A.B. Zamolodchikov,
Teor. Mat. Fiz. {\bf 65} (1985) 1205.}
%\REF\bpz{A.A. Belavin, A.M. Polyakov and A.B. Zamolodchikov, {  Nucl.
%Phys.} {\bf B241} (1984) 333.}
%\REF\bow{P. Bouwknegt and K. Schoutens, CERN preprint CERN-TH.6583/92,
%to appear in Physics Reports.}
\REF\he{C.M. Hull, \pl\ {\bf 240B} (1990) 110.}
\REF\hee{C.M. Hull,  \np\ {\bf 353
B} (1991) 707.}
\REF\heee{C.M. Hull, \pl\ {\bf 259B} (1991) 68.}
%\REF\vanbrs{K. Schoutens, A.
%Sevrin and P. van Nieuwenhuizen, \cmp\ {\bf 124} (1989) 87.}
\REF\wanoms{C.M. Hull, \np\ {\bf B367} (1991) 731;   \pl\ {\bf 259B}
(1991) 68; QMW preprint QMW/PH/91/14 (1991);
  K. Schoutens, A. Sevrin and
P. van Nieuwenhuizen, \np\ {\bf B364} (1991)  584 and {\bf B371} (1992) 315;
in proceedings of 1991 Miami Workshop (Plenum,
New York, (1991)).}
\REF\wstr{C.N. Pope, L.J. Romans and
K.S. Stelle, \pl\ {\bf 268B} (1991) 167 and {\bf 269B}
(1991) 287.}
\REF\thierry{J. Thierry-Mieg, \pl\ {\bf 197B} (1987) 368.}
\REF\wstrbil{A. Bilal and J.-L. Gervais, \np\ {\bf B326}
(1989) 222.}
\REF\romans{L.J. Romans, \np\ {\bf B352} 829.}
\REF\das{S.R. Das, A. Dhar and S.K. Rama, {Mod. Phys. Lett.}
{\bf A6} (1991) 3055; {Int. J. Mod. Phys.} {\bf A7} (1992) 2295;
S.K. Rama, {
Mod.\ Phys.\ Lett.}\ {\bf A6}
(1991) 3531.}
 \REF\pzz{C.N. Pope, L.J. Romans, E. Sezgin and K.S. Stelle,
{  Phys. Lett.} {\bf B274} (1992) 298.}
\REF\paa{H. Lu, C.N. Pope, S. Schrans and K.W. Xu,  {  Nucl.
Phys.} {\bf B385} (1992) 99.}
\REF\pbb{H. Lu, C.N. Pope, S. Schrans and X.J. Wang,  {  Nucl.
Phys.} {\bf B379} (1992) 47.}
\REF\pcc{H. Lu, B.E.W. Nilsson, C.N. Pope, K.S. Stelle and P.C. West,
``The low-level spectrum of the $W_3$ string,'' preprint CTP TAMU-64/92,
hep-th/9212017.}
 \REF\ewrwe{C.N. Pope, E. Sezgin, K.S. Stelle and X.J. Wang, \pl\
{\bf B299} (1993) 247.}
\REF\westsolo{P.C. West, ``On the spectrum, no ghost theorem and modular
invariance of $W_3$ strings,'' preprint KCL-TH-92-7, hep-th/9212016.}
\REF\pff{H. Lu, C.N. Pope, S. Schrans and X.J. Wang,  ``The interacting
$W_3$ string,'' preprint CTP TAMU-86/92, KUL-TF-92/43, hep-th/9212117.}
 \REF\pjj{M. Freeman and P. West, ``$W_3$ string scattering,''
preprint, KCL-TH-92-4, hep-th/9210134.}
 \REF\pooop{H. Lu, C.N. Pope, S. Schrans and X.J. Wang,
``On the spectrum
and scattering of $W_3$ strings,'' preprint CTP TAMU-4/93, KUL-TF-93/2,
hep-th/9301099.}
\REF\west{M.D. Freeman and P.C. West, ``The covariant scattering and
cohomology of $W_3$ strings,'' preprint, KCL-TH-93-2, hep-th/9302114.}
\REF\lat{H. Lu, C.N. Pope and X.J. Wang,
\lq On Higher-spin Generalisations of String Theory', preprint
CTP TAMU--22/93, (1993)
 hep-th/9304115.}
\REF\bergo{E. Bergshoeff,    A. Sevrin, and X. Shen,
\pl\ {\bf B296} (1992) 95.}
\REF\bergs{E. Bergshoeff, H.J. Boonstra, M. de Roo, S. Panda and A.
Sevrin,
``On the BRST operator of $W$ strings,'' preprint,
UG-2/93, UCB-PTH-93/05,LBL-33737.}
\REF\bata{I.A. Batalin and G.A. Vilkovisky, \pl\ {\bf 102B} (1981) 27
and \pr\ {\bf D28} (1983) 2567.}
\REF\dotso{V. Dotsenko and V. Fateev, \np\ {\bf B240} (1984) 312;
{\bf B251} (1985) 691.}
\REF\felder{G. Felder, \np\ {\bf B317} (1989) 215.}
\REF\thorn{C. Thorn, \np\ {\bf B286} (1987) 61; Phys. Rep.
{\bf 175} (1989) 1.}
\REF\thornkac{C. Thorn, \np\ {\bf B248} (1984) 551.}

\def\np{Nucl. Phys.}
\def\pl{Phys. Lett.}
\def\pr{Phys. Rev.}

\chapter{Introduction}

There are an infinite number of extended conformal algebras --
super-conformal algebras, W-algebras, topological  conformal algebras,
fractional supersymmetry algebras etc -- and it seems that
corresponding to each of these  one has the possibility of constructing
a generalisation of string theory. An important question  is whether
these are really fundamentally new string theories, or whether they are
equivalent to ones that are already known. The first step in the
construction of such theories is to promote the semi-local extended
conformal symmetry of some matter system to a fully local symmetry by
coupling to world-sheet gauge fields (supergravity, W-gravity,
topological gravity etc). For the   case   of the $W_3$ algebra [\zam],
  the corresponding W-gravity theory was constructed in [\he] and it
was subsequently shown that this construction will  work for {\it any}
 extended  conformal algebra [\hee,\heee].  Quantising this system
defines a generalised string theory (superstring, W-string, fractional
superstring etc)
  and the string theory will be critical if the matter system is chosen
in such a way that   all  anomalies in the local gauge symmetries are
cancelled by ghost contributions.   For non-linearly realised
symmetries of the type that arise in W-gravity, the cancellation of
anomalies is much more delicate than in the  usual string and
superstring theories [\wanoms]. For the $W_3$ algebra, it was finally
shown [\wstr] that all anomalies will cancel if  the matter system  is
a $c=100$ realisation of $W_3$, and the  physical states correspond to
the cohomology classes of the BRST operator constructed in [\thierry],
as anticipated in [\wstrbil]. A $c=100$ realisation of $W_3$ can be
constructed by taking  any
conformal field theory with $c= 25 \ha$ and adding to
it a single free boson with background charge [\romans],
and all $c=100$
realisations that are known take this form.

A similar construction works for many other extended conformal algebras.
For example, to construct a critical $W_N$ string theory, it is
necessary  to find a matter theory with $W_N$ symmetry
whose central charge takes the
  critical value
  $c_N^*=2(N-1)(2N^2+2N+1)$.
One way of attaining such a realisation is given by taking an arbitrary
conformal
 field theory with central charge $$\tilde c= 26 - \left(1 - {6  \over
N(N+1)} \right) \eqn\ceffn$$ and adding $N-2$    free bosons $ \phi ^a$
($a=1, \dots ,N-2$) with background charge. Remarkably,
$\tilde c
 = 26- c_{N-min} $
 where  $ c_{N-min}$ is the central charge of the $N$'th minimal model.
 In the last year or so, a large amount of effort has been devoted to
the  spectra and interactions of such theories [\pzz -\lat]; in the
case of the $W_3$ string, the spectrum is now known explicitly for the
first few levels, suggesting some interesting conjectures which have so
far not been proven (beyond the first few levels).
 It appears that in each case the physical states can be constructed
from the  primary fields of the effective theory with central charge
$\tilde c$ whose conformal dimensions are $$ \Delta = 1- h^N_{r,s}
\eqn\fhglasakfjg$$ where
$h^N_{r,s}$ are the weights of the $N$'th minimal model.
This suggests a close relation
 between the $W_N$ string and the $N$'th minimal model [\das-\lat],
and there are generalisations of this construction which
correspond to minimal
series for other W-algebras [\pbb].\foot{Note that
the  values of the central charges and  intercepts given by
\ceffn,\fhglasakfjg\ are the only ones for which it is possible to have
a unitary conformal field theory with $25<c<26$ [\thornkac].} These
results suggest that there might be a sense in which such W-strings
could be regarded as ordinary $c=26$ strings in which part of the
matter sector is taken to be a minimal model [\lat], but it has not
been clear how this could be done.

The purpose of this paper is to show that it is indeed the case
that $W$-strings can be thought of as ordinary strings with
minimal-model  sectors, and to investigate the new realisations of
minimal models that arise  in these theories. The
realisation of the Ising model that occurs in the $W_3$ string will be
analysed in detail, drawing heavily on the results for the spectrum and
interactions of the $W_3$ string derived in [\pzz -\lat]; indeed,  some
of what follows is implicit in the work of [\pzz -\lat], but
the reformulation given here considerably
clarifies
the structure of W-strings and will be used to analyse the complete
spectrum of the theory.
A key ingredient is the
fact that, after a field redefinition,  the structure of
$W_3$ strings based on the $W_3$ repersentation of [\romans]
simplifies dramatically and the
BRST charge for $W_3$ strings can be written as $\hat Q=Q +Q_0$
where $Q_0$ is the standard bosonic string BRST operator and
$Q$ acts only on the ghost system for the spin-three symmetry and
the single free boson [\pooop,\bergs].
The result of the analysis indicates that, at
least at tree level,  the  $W_N$ strings that have been
constructed so far can be viewed as
ordinary  string theories in which the matter sector
consists of a minimal model plus an arbitrary conformal field theory
with central charge \ceffn.

This can be understood as follows. The $W_3$ string is constructed from
a matter system $M$ with $c=100$ that has $W_3$ symmetry, together with
the usual $c=-26$ $b,c$ ghost system, and a $c=-74$ W-ghost system $\b,
\g$. These fields can be re-organised into the $bc$ ghosts  plus the
$c=26$ system $(M,\b,\g)$, so that at tree level the theory is a
conventional string  which happens to have a particular type of matter
system with extra symmetry (in the same sense in which a superstring
can be regarded as a conventional ($c=26$) string in which the matter
sector consists of a $c=11$ commuting $\b \g$ system, plus a $c=15 $
conformal field theory which has extra super-conformal symmetry).
However,   the only $c=100$ conformal field theories with    $W_3$
symmetry that
are known  are constructed from a $c=25\ha$ conformal matter system $X$,
which need have no extra symmetry, plus a single boson $\f$ with
background charge and $c=74 \ha$. The $(\f,\b,\g)$ system has $c= \ha$
and will be shown in this paper to be essentially the Ising model at
its critical point. The spin-three symmetry acts entirely in the
 $(\f,\b,\g)$ system, and no restrictions are imposed on the $c= 25\ha$
conformal matter system  $X$. Thus the reducibility of the theory is a
consequence of the  reducibility of the particular representation of
the $c=100$ {} $W_3$ algebra that is used, and would not hold for a
matter system that did not reduce in this way. Even for the reducible
cases, it may often be convenient to  consider such models from the
viewpoint of
 W-strings, so that the extra symmetry that arises for these particular
cases of the bosonic string is more manifest and can be used to
organise the analysis. It remains to be seen whether this view-point
can be
applied at higher genus   or whether, when moduli are taken into
account, the theories are  really distinct.

\chapter{Free Field Realisation of the Ising Model}

Consider a conformal theory consisting of a free boson $\f$
and a   ghost/anti-ghost  system $ \b , \g$ where $ \b , \g$
are anti-commuting and have spins $3$ and $-2$, respectively.
The stress-tensor is
 $$T=T_{\f}+T_{\b \g}\eqn\ttis$$
where
$$T_{\f}=- {1\over 2} \pa \f \pa \f - q \pa ^2 \f,
\qquad
T_{\b \g} = -3 \b \pa \g -2 \pa \b \g
\eqn\tisis$$
and $q$ is a background charge.
The central charge of the $ \b , \g$ system is $c_{\b \g} =-74$,
while the
scalar field central charge is
$c_ \f=1+12 q^2$. Choosing
$q^2= {49 \over 8}
$
gives $c_\f= 74 {1 \over 2}$ so that the total central charge of the
$\f,\b,\g$ system is ${1 \over 2}$.
This free field theory can be quantised in the conventional way
 to give
  a
non-unitary reducible conformal field theory with $c= \ha$.
The theory   has a conserved primary spin-three current
given by
$$\eqalign{
&W=  W^\f +W^{\b \g}
\cr &
W^\f=
{1 \over 3} (\pa \f)^3 +q \pa ^2 \f \pa \f +\left( {q^2 \over 3}-
{5 \over 4} \right)\pa ^3 \f
\cr &
W^{\b \g}=
 \pa Y \g + 2 Y \pa \g,
\qquad Y\equiv \ha ( 3\pf \b + q \pa \b)
\cr}
\eqn\etert$$
 The
commutator $[W(z),W(w)]$ can be written in terms of
$T$ and a spin-four current
$W^{(4)}$, and $W^{(4)}$ can be written as a sum of
terms each of which
involves either
$W$ or $\b$. Indeed, the
currents $T,W,\b$
generate a closed W-algebra with field-dependent
structure functions.
Alternatively, one can regard $W^{(4)}$ as an
independent new current
and
closing the algebra leads to an infinite set of currents
$W^{(n)}= {1 \over n}
(\pf)^n + \dots$ for $n=2,3,4,\dots$, with
$W^{(2)}=T$ and $W^{(3)}=W$, and  these satisfy a $W_\infty$
algebra with $c= \ha$.

Consider the BRST operator
$$\eqalign{
\QQ &=   \int dz \, \g \left[ W^\f + {1 \over 2} W^{\b \g} \right]
\cr
&=\int dz \,\g\left[
{1 \over 3} (\pa \f)^3 +2q \pa ^2 \f \pa \f + \left({ q^2\over 3}
-{5 \over 4}\right) \pa ^3 \f
+ {3\over 2} \pf \b \pa \g
+  {q\over 2} \pa \b \pa \g
\right]
\cr}
\eqn\qqis$$
This is nilpotent if $q^2=49/8$,
$
\QQ ^2 =0$, and commutes with the stress-tensor $T$,
the spin-three current $W$ and hence with all of the $W_\infty$
currents
$W^{(n)}$. We shall define the physical states of our model to
be the
cohomology classes of $\QQ$.  The stress-tensor is BRST
non-trivial,
but $W$ is BRST exact: $$ \{ \QQ, \b \} =W \eqn\wtriv$$ as
are all of
the higher-spin currents $W^{(n)}$ for $n \ge 3$.
Thus the W-sector of
the theory is \lq topological', while the ordinary sector is not
(correlation functions are independent of any background
 spin-three
gauge field $B$ but not of the metric etc).

This theory can be thought of as arising from a model with
Lagrangian  $$L= \pf \bar \pa \f + B\tilde W, \qquad \tilde W=
{1 \over 3}( \pf)^3 \eqn\lis$$
  corresponding to a free boson  subject to the constraint
$\tilde W=0$ imposed by the Lagrange multiplier $B$, which can
be viewed as a spin-three gauge field for the W-gravity based
on the W-algebra with the single generator $\tilde W$ (which
closes since $[\tilde W(z),\tilde W(w)]$ can be written in
terms of $\tilde W$, with field-dependent structure functions).
Choosing the gauge  $B=0$ and introducing ghosts $\b, \g$ leads
to the naive (\lq classical') BRST current  ${1 \over 3} (\pa
\f)^3  + {3\over 2} \pf \b \pa \g $ [\bergs]. A nilpotent
quantum BRST operator is obtained by adding  quantum corections
to this to obtain \qqis\ [\bergs], and  including a background
charge $q$ in the stress tensor ensures that $[\QQ,T]=0$
provided $q^2=49/8$. These quantum corrections can be
systematically derived as in  [\bergo] using the methods of
Batalin and Vilkovisky [\bata].

As $[\QQ , T]=0$,   the cohomology classes can be organised
into  representations of the Virasoro algebra  so that it is
sufficient to restrict attention to those classes represented
by  primary fields, as all other ones will be    represented
by   descendents of these. (In fact, they can be arranged into
representations of the $W_\infty$ algebra, as will be discussed
in the next section.)

The cohomology classes represented by primary fields all
have conformal dimension $\dd$ given by $0, \ha$ or ${1
\over 16}$.
The operators
$$
V_0 = \pa \g \g e^{- {8\over  7}q \f}
,
\qquad
 V_ {{1 \over 16 }}
= \pa \g \g e^{-  q \f}
,
\qquad
V_\ha  =   \g e^{- {4\over  7}q \f}
\eqn\vis$$
all represent non-trivial cohomology classes and have
conformal dimension   $0, {1 \over 16}$ and $\ha$  respectively.
These will be identified with the identity operator
  {\bf 1},    the spin operator $\sigma$  and the
  energy operator $\varepsilon$
of the Ising
model.
The operator $V_ {{1 \over 16 }} $ is self-conjugate,
while the conjugate of $V_0$ is
$\bar V_0 = \pa \g \g e^{- {6\over  7}q \f}$, since   $a_0^\dagger
=a_0-2iq$ [\pcc].

The  picture-changing operator
$$
P= [Q, \f]=
- {19 \over 24} \pa ^2 \g -{3 \over 2} \b \pa \g \g -
\pf \pf \g + q \pf \pa \g
\eqn\pis$$
  clearly satisfies $\{Q, P\} =0$ and has dimension $0$.
As usual, this is not regarded as being
 BRST trivial   as $\f $ is not a proper
conformal field.
Thus given a primary cohomology class represented by $V$,
there is
another one of the same conformal dimension represented by
$PV=:PV:$,
and the classes can be organised into  doublets
$\{ V,PV \}$
(note that $P^2V$ is BRST trivial).

In addition, there are cohomology classes represented by
non-local operators; if a
field $U(z)$  satisfies
$$[Q,U]= \pa Y
\eqn\dec$$ for some $Y$, then
$S_U=\int dz U$ will satisfy $[\QQ, S_U]=0$ and so will represent
a (possibly trivial) cohomology class. Of particular interest
are such classes for  which $U$ has dimension one, so that
the corresponding
$S_U$ are  well-defined and have dimension zero; these   are the
 screening
operators of the theory and play
a role similar to the  screening
operators in the Coulomb gas representation of the minimal models
[\dotso,\felder].

The standard screening operators formed from the
spin-one Virasoro primary fields $e^{iq_{\pm}\f} $
$$A_\pm = \oint dz e^{iq _\pm \f},
\qquad
q_\pm (q_\pm -2q^2)+2=0
\eqn\qpm$$
do not commute with the BRST operator and so do not play a role here;
the screening operators
of this theory all take the form $\oint dz K(\b,\g,\pf) e^{-p \f}$
for some momentum $p$ and some operator $K$ constructed from
$\pf ,\b$ and $\g$.
 The
 basic screening operators of the theory are
[\west] $$S= \oint dz \, \b e^{  {2\over  7}q \f}
\eqn\sis$$
and
$$R= \oint dz \,\g e^{ - {6\over  7}q \f}, \qquad \bar R=
\oint dz \, \g e^{ -
{8\over  7}q \f} \eqn\spis$$
There are in fact an infinite number of screening operators
$S(n), \bar
S(n)$ ($n=0, \pm 1 , \pm 2, \dots$) with $S(0)=R, \bar S(0)= \bar R,
S(1)=S$, which will be constructed presently. Given any local physical
operator $V(z)$,
 new physical operators can be constructed  by acting with the
screening charges $S(n),\bar S(n)$ and the picture changing operator
$P$. For a set of screening charges charges $S_i= \oint dz U_i(z)$, the
product is defined by its action on $V$:
 $$ (\prod_{i=1}^n S_i V)(z) = \oint _{C_1} dw_1  \oint _{C_2} dw_2
\dots  \oint _{C_n} dw_n U(w_1) U(w_2) \dots U(w_n) V(z) \eqn\svis$$
where the contour $C_n$ surrounds the point $z$, the contour $C_{i} $
surrounds  $C_{i+1} $ and all   the contours are taken to touch each
other in precisely one point, as in [\felder,\west].
 If $U_i$ contains a factor $exp(- a \f)$ and $V$ contains a factor
$exp(-b \f)$, then $S_iV$ will only be a well-defined local operator if
the product of the \lq momenta' $ab$ is an integer, and there are
similar restrictions   on the momenta of the $U_i$ and $V$ for the
construction \svis\ to be well-defined [\west]. When a new operator
obtained in this way is well-defined and non-zero, it will represent a
non-trivial  BRST cohomology class if the original operator $V$
did, and will have the same conformal dimension as the original
operator,
but different momentum and ghost number in general.

Acting with $S$ and $P$ on
 the three operators \vis, the only local operators that can be
constructed are defined by [\west]
$$\eqalign{&\bar V(0,n)=SPV(0,n), \qquad
V(0,n)= S^3P \bar V(0,n-1), \qquad V(0,0)=V_0 \cr & V(\si ,n)=
S^2P V(\si,n-1), \qquad
V(\si,0)=V_\si
 \cr & V(\ha, n)=S^3P\bar V(\ha, n-1),\qquad \bar V(\ha,
n)=SPV(\ha, n),\qquad V(\ha, 0)=V_{\ha} \cr } \eqn\vnis$$ plus the
picture-changed partners given by acting with $P$ on each of these.
The operator
 $V(\Delta,n)$ has conformal dimension $\Delta$. This series
can be extended to negative $n$ by acting with $R,\bar R$:
$$   \eqalign{& V(0,n-1) \propto RPV(0,n-1)
\cr
&V(\si,n-2) \propto RP V(\si,n)
\cr &
 V(\ha , n-1) \propto RPV(\ha , n-1)
\cr}
\eqn\vnles$$
For example, $V(0,-1)$ is a new physical
vertex operator of conformal weight zero and ghost number four
given by
$\g \pa \g \pa ^2 \g \pa ^7 \g exp(-16q\f /7)$.
The operators $V(0,n),V(\si,n)V(\ha,n)$ have ghost numbers given
respectively by $2-2n,2-n,1-2n$ and momenta given by $p=iq k/7$ with
$k$ given by $8-8n,7-4n,4-8n$ respectively. Thus there are operators
for arbitrarily large positive or negative ghost number and imaginary
momentum.

The operator $V(0,1)$ is the identity operator $V(0,1)={\bf 1}$,
and so the  whole family of dimension zero operators $V(0,n),\bar
V(0,n)$ are obtained by acting on ${\bf 1}$ with $P$ and screening
operators. The operator $x= \bar V(0,1)$ also has dimension zero and
ghost-number zero, and acts on $\bar V_0= \bar V(0,0)$ to give $x \bar
V_0=V_0$.  The
product $x^n$ will be well-defined if $n=4m$ or $n=4m+1$ and the
operators $1, x^4$ generate the ground ring of the theory [\ewrwe].
(The  operator $V(\si , 2)$ also has
  ghost number zero.)

Given any screening operator $S= \oint dz U$,  the BRST variation of
$U$ is $[Q,U]= \pa V$ for some $V$. However, this $V$ must have
dimension zero and be BRST-invariant, and conversely given any of the
dimension-zero BRST-invariant operators $V(0,n), \bar V(0,n)$ defined
by \vnis,\vnles, one can solve $$[Q,U(n)]= \pa V(0,n), \qquad
 [Q,\bar U(n)]= \pa \bar V(0,n) \eqn\ququ$$ to obtain dimension-one
operators $U(n),\bar U(n)$. The screening charges $S(n),\bar S(n)$  are
defined as the integrals of the operators $U(n),\bar U(n)$, which are
determined by the descent equation \ququ. Acting on the set of
operators \vis\ with these screening operators gives no further
operators; note in particular that $S(2)V(0,0)$ gives the identity
operator, $1$. Note also that not all of these screening operators are
non-trivial e.g. $U(1)$ is trivial since $V(0,1)=1$.

Corresponding to each of the three basic operators \vis, there is an
infinite set of \lq screened and picture-changed' versions of the same
physical operator, all of which have the same conformal dimension, but
have different ghost-numbers and different discrete values of the
momenta in general. As in the Coulomb gas models, these are to be
interpreted as physically equivalent versions of the three original
operators, and the different screened versions are needed to construct
amplitudes.
 All the operators are generated by the action of $S,R,\bar R , P$ on
three basic operators, which can be taken to be those given by \vis,
although for some purposes it is more natural to represent
the dimension zero operators by ${\bf 1}$. As
a result of \vnis,\vnles, the action of a string of $S$'s and $P$'s,
when defined, can be inverted by the action of appropriate $R$'s and
$P$'s, and vice versa, and  this gives a strong indication    that
these should be all the independent local operators and screening
operators, since given any  vertex operator of ghost number $N$, acting
on it with a suitable string of $S$'s, $R$'s and $P$'s should bring it
to one of low ghost number, and this
 should be one of the ones already found. Then we can invert the action
of   the $S$'s, $R$'s and $P$'s by a further such string to regain the
original operator, so the original operator can be written in terms of
the action of $S$'s, $R$'s and $P$'s
on one of the operators in \vis, and so must be one of the operators
already constructed.
To complete this argument, it would be necessary to show that
there is no  operator that can't be brought  to one of low ghost
number by the action  of the screening and picture operators, and to
complete the classification of operators of low ghost number begun in
[\das-\lat].

In the next section, further evidence will be given that  the operators
of the theory are all obtained from basic operators using screening and
picture operators, so that the spectrum should be that of the Ising
model (regarding different screened and picture-changed versions of a
given operator as being physically equivalent). In
section 4 the correlation functions will be considered, and shown to be
consistent with the assertion that this constrained free field theory
is equivalent to the Ising model.

\chapter{BRST Cohomology}

In this section,  the BRST cohomology of the $(\f,\b,\g)$
theory  will be
analysed; further details will be given elsewhere. As all the
$W_\infty$
generators are $Q$-closed, $[\QQ,W^{(n)}]=0$,   the cohomology classes
of physical operators can be organised into  representations of the
$W_\infty$ algebra, so that it is sufficient to restrict attention to
those classes represented by $W_\infty$-primary fields, as all other
ones will be    represented by   descendents of these. Such primary
fields correspond to states $|\Psi> $ that  are annihilated by the
modes $W_{n}^{(m)}$ for $n >0,m \ge 2$ and are eigenstates of
$W_{0}^{(m)}$.  Since for spins $m>2$ the $W_\infty$ generators are
exact, $W_{0}^{(m)}=\{ \QQ , V_{0}^{(m)} \}$  for some $ V_{0}^{(m)} $,
it follows that any $\QQ$-closed state   with non-zero $W_0^{(m)}$
eigenvalue $ \lambda^{(m)}$ for $m>2$ must be BRST trivial:
$$ W_0  ^{(m)}|\Psi>= \lambda ^{(m)} |\Psi> \quad
\Rightarrow
\quad
 |\Psi>= {1 \over \lambda ^{(m)}}\QQ V ^{(m)}|\Psi>
\eqn\coho$$
Thus
for a state $|\Psi>$ to represent a
non-trivial cohomology class,
it must satisfy
$W_0^{(m)}|\Psi>=0$ for all $m>2$, but the
Virasoro weight (the eigenvalue of $L_0$) can be non-zero.
Furthermore,  the descendents obtained by acting on such a
physical  highest
weight state with polynomials in $W_{-n}^{(m)}$ for $m>2$ are all BRST
trivial, so that the classes can be represented by
 $W_\infty$ highest-weight states or by the Virasoro descendents
obtained by acting with polynomials in $L_{-n}$. The representatives of
the classes can also be chosen to have definite  ghost-number and
momentum $p$ (the eigenvalue of $a_0$, where  $a_n$ are the  modes of $
i\pf$). The spin-three constraint $ W_0   |\Psi>=0$  then implies
that the momentum $p$
  must satisfy a cubic equation, so that for, a fixed level and ghost
number etc,
 the momentum must be frozen to one of three discrete values, and
for physical states this turns out to be imaginary and quantised in
units of $q/7$: $p=iq N/7$ for some integer $N$.

Modes are defined as usual by $$\eqalign{& \b = \sum _n \b _n z^{-n-3},
\qquad \g = \sum _n \g _n z^{-n+2}, \cr &
T= \sum L_n z^{-n-2}, \qquad W = \sum _n
W_n z^{-n-3}, \cr &
i\pf = \sum _n a_n z^{-n-1}, \qquad P= \sum _n P_n z^{-n}
\cr }
\eqn\modes$$
Note that since each of the operators $\b_n,\g_n$ is nilpotent,
there is an associated cohomology for each and
the interplay between these, the
de Rham cohomology on the world-sheet
 and the $Q$-cohomology leads to a series of
descent equations that play an interesting part in the analysis.

The
state space of the theory is spanned by the states
$$a_{-n_1}a_{-n_2}...a_{-n_r}
\b_{-m_1}\b_{-m_2}...\b _{-m_s}
\g_{-p_1}\g_{-p_2}...\g_{-p_t}|p>
\eqn\state$$
where $|p>$ is the Fock vacuum with
 momentum $p$ and  zero ghost-number satisfying
$$\eqalign{&
a_{n}|p>=\b_n |p>=
\g_n|p>=L_n|p>=W_n^{(m)}|p>=0
\cr &
  (a_0-p)|p>=\b_0|p>=(L_0-\Delta)|p>=0
\cr}\eqn\vac$$
for $n>0$.
For generic momenta $p$, we can make the following change of basis for
$n>0$:
$$\eqalign{ &a_{-n} \rightarrow \hat a_{-n} \equiv L_{-n}=
a_{-n}P(n) + \dots, \cr &
\g_{-n} \rightarrow \hat \g_{-n} \equiv [Q,a_{-n}]
=n(\pa P)_{-n}= G(n) \g_{-n}+ \dots,
\cr & \g_0 \rightarrow \hat \g _0 \equiv P_0
\cr}
\eqn\chbas$$
where
$$\eqalign{P(n)&=(a_0+i(n+1)q), \cr
G(n)&=(n+2)\left[ a_0^2+(n+1)\left(i a_0q-{19 \over 24}n\right)\right]
 +.....
\cr}
\eqn\jac$$
(note that $G(n)$ contains ghost-dependent terms).
For certain special discrete values of the momentum $p$,  $P(n)$ or
$G(n)$ will vanish for some $n$ and for   those momenta
this change of basis is singular and     extra discrete states can
arise. However, for any value of $p$ not lying in this discrete set,  a
basis of states with that momentum is given by those of the form
 $$\hat a_{-n_1}\hat a_{-n_2}...\hat a_{-n_r}
\b_{-m_1}\b_{-m_2}...\b _{-m_s}
\hat \g_{-p_1}\hat \g_{-p_2}...\hat \g_{-p_t}|p>
\eqn\stateh$$
where $|p>$ is now taken to satisfy \vac\ with
$a,\g$ replaced by $\hat a, \hat \g$.
The advantage of this basis is that $\hat a_n$ and $\hat \g _n$
(anti-)commute with $Q$.
There will also be certain discrete values of the momenta at which
W-algebra singular vectors occur, as we shall see below.
However, first we will deal with the case of generic momenta at which
the change of basis is non-singular and there are no
W-algebra singular vectors.
For such generic momenta,   $\hat \g_{-n}$ is
BRST exact for $n>0$
and
$\{ Q, \b_{-n} \} = W_{-n}$  is a non-singular operator,
and it is then straightforward to argue (using e.g. the methods of
[\thorn]) that the BRST classes can   be represented by states
involving no $\b_{-n},\hat \g_{-n}$ oscillators for $n>0$.
This leaves the states $|p>$
and the descendents  obtained by acting on these with
the $\hat a_n= L_{-n}$ and the picture-changing operator
$\hat \g _0=P_0$. Acting with $P_0$
more than once gives a BRST trivial state,
so  the physical states fit into picture-changed doublets
$\{ |\Psi>, P|\Psi>\}$.
These descendents will represent  non-trivial
$Q$-cohomology classes if and only
if the state $|p>$ does. The  constraint $W_0|p>=0$
gives a cubic equation for $p$ whose   solutions are
$p=iq, 6iq/7, 8iq/7$ and these are  the states $|p>$ given
by acting on the $SL(2)$ invariant vacuum $|\Omega>=
\b _{-1} \b_{-2}|0>$ with
$  V_\si, V_0,\bar V_0$ and their $P$-doubles. They are   BRST
invariant and represent non-trivial cohomology classes.

All other physical states must occur at those discrete values
of the momentum at which either the change of basis is singular, or
there are singular vectors.
(Note that the values of these
discrete physical momenta   depend on the
ghost number in general.)
 In each case the extra discrete states  fit into
multiplets consisting of a highest weight state, together with the
descendents obtained by  acting with the $L_{-n}$, plus the doubles
obtained by acting once with  $P_0$.

The Jacobian for the change of basis $a_{-n}
\rightarrow L_{-n}$ will
vanish when $P(n)$ does, \ie\ for momenta given by
$p=-(n+1)iq$ for some integer $n$. For these momenta, there are
  Virasoro singular vectors and so there are states
in the scalar Fock space  that are annihilated  by linear combinations
of the Virasoro generators, so that the Verma module generated by the
$L_{-n}$ does not fill the Fock space. For
momentum $p=-(n+1)iq$, this means that the argument
eliminating the oscillator $a_{-n}$ breaks down and there is the
possibility of
extra physical states involving that oscillator.

The next possibility is that the change of basis
$\g  \rightarrow \hat \g$ is singular, will occur for momenta
and ghost numbers at which $G(n)$ vanishes for some $n$.
This will occur if there are states that are annihilated by
some function of the $\hat \g_{-n}$ in a non-trivial way.
 For example, given a $Q$-closed state $|\Psi >$
satisfying $\hat \g_{-n}|\Psi>=0$, the state
$a_{-n}|\Psi>$ will also be $Q$-closed.
More generally, given a state $|\Psi>$ which is annihilated by
some function $f$ of the $\hat \g$ given by $f= \hat \g_{-n}+...$
for some $n$ and which is not of the form
$|\Psi> =\hat \g _{-n}|\Theta>+...$, there should be a physical state
of the form $a_{-n}|\Psi>+...$
If $G(n) \ne 0$, then $a_{-n}|\Psi>=Q|\Theta>+...$
where $|\Theta>=\ha G(n)^{-1} a_{-n}^2 \b _n |\Psi>+...$
so such states can only be non-trivial when $G(n)=0$, and reflect the
singularity of the change of basis.

Finally,
  there there are the momenta  at
which the argument eliminating the ghost, anti-ghost or scalar modes
might break down, due to the presence of singular vectors of the
$W$-algebra. Consider, then, the momenta
 at which $W_\infty$ singular vectors occur. These correspond to
states in the Fock-space which are annihilated by polynomials in the
$W_\infty$ generators $W_{-n}^{(m)}$. Consider first the simple case of
momenta at which there is a state $|\Psi >$ annihilated by one of the
modes of the spin-three current,  $W_{-n}|\Psi>=0$. Then the argument
leading to the elimination of the  corresponding  anti-ghost $\b_{-n}$
breaks down and the state  $\b_{-n}|\Psi>$ represents a new non-trivial
cohomology class. However, since $\b_{-n}$ is nilpotent, it is not
necessary for $|\Psi> $ to be annihilated by $Q$ for this to work; it
is sufficient that the state satisfy the descent equation
$$Q|\Psi>=\b_{-n}|\Phi>
\eqn\deco$$
for some $|\Phi>$,
as this implies
$ Q \b_{-n}|\Psi> = W_{-n}|\Psi>$
which vanishes.

The general situation is that in which
a state is annihilated by some polynomial
$f(W_{-n},\b_{-m},\g_{-r},L_{-s})$; nothing extra is gained by including
the modes of the higher spin currents $W^{(s)}$, as these can all be
rewritten in terms of multiple commutators of the modes of $W$, as
$W^{(s)}_n$ arises in the commutator $[W_m^{(3)},W_{n-m}^{(s-1)}]$. For
a state $|\Psi>$ annihilated by a polynomial $f=W_{-n}+...$, it is
expected that there should be a physical state of the form
$\b_{-n}|\Psi>+...$ and in general such states can be constructed
perturbatively.

As a first example,   the (non-physical) state $|p>$ with $p=4iq/7$
satisfies $$(W_{-1}-\lambda L_{-1})|p>=0
\eqn\wcon$$
with $\lambda = -12q/7$, suggesting that we consider the state
 $\b _{-1}|4iq/7> $; this is in fact physical without
any further modification as
$Q|4iq/7>= \lambda \g _0 |4iq/7>  $.
This gives the physical state $\b _{-1}|4iq/7>=V_\ha |\Omega> $.

As another example, the (non-physical) zero-momentum state $|0>$
is annihilated by an operator of  ghost-number $-1$, as
it
satisfies
$$(\b_{-1} W_{-2}-\b_{-2}W_{-1}-12 q \b_{-1}\b_{-2}\g_0)|0>=0
\eqn\ststst$$
This implies that there is a physical state at ghost-number $-2$
given by
$\b_{-1}\b_{-2}|0>$, which is precisely the $SL(2)$ invariant vacuum.
The constraint on the ghost-number zero state $|0> $ appears to have
non-trivial ghost-dependence, but the situation becomes clarified
by rewriting
\ststst\ in the form
$$\left(W_{-2} + \ha W_{-1}L_{-1} -{15q \over 4}
L_{-1}^2\right)\b_{-1}|0>=0
\eqn\zzz$$
which is  a conventional (ghost-independent)
constraint on the state $\b_{-1}|0>$ of ghost number $-1$.
{}From \zzz, we see that  there should be
a physical state of the form $\b_{-2}\b_{-1}|0>+...$,
 and again it turns out that no further corrections are needed,
so that $|\Omega>=\b_{-2}\b_{-1}|0>$ is a physical state, the SL(2)
invariant vacuum.
More generally,  given a W-algebra singular vector
 of ghost number $N$, there should be a physical
state at ghost number $N-1$
and it appears that the physical states constructed
in the last section can
be viewed as arising in this way.

In this section it has been seen that extra physical states arise only
from singular vectors of the $W$-algebra, or from singular vectors of
the  $\hat \g _{-n}$ operators. From the last section, it was seen
 that the same physical states can be constructed using screening
operators, so that there should be a screening charge construction of
the W-algebra singular vectors, generalising the screening charge or
vertex operator  construction of Virasoro singular vectors
[\thornkac]. To complete the analysis requires a classification of such
singular vectors or the corresponding screening operators and work in
that direction is currently in progress. However, the picture that is
emerging supports the conjecture that all the physical operators are
obtained by acting on the three basic operators \vis\ with screening
and picture-changing  operators, so that there are just three physical
operators in the theory. Note that one might also regard the theory as
the Ising model tensored with a $c=0$ conformal theory whose spectrum
consists of an infinite number of discrete ground states of zero
conformal weight, giving an infinite number of copies of each of the
three Ising operators, and that the Ising model emerges on identifying
the copies of each Ising operator, \ie\ by factoring out the discrete
theory.
It is important to check that identifying the different
operators of a given spin gives a sensible well-defined theory; in
the next section we will check that the correlation functions are
well-defined and agree with those of the Ising model.

\chapter{Correlation Functions}

Since the model in question is a free field theory, it is
straightforward to calculate correlation functions. Consider first the
partition function. Choosing one representative for each of the three
physical operators,
 the analysis of [\westsolo] suggests that the partition function of
the $\f,\b,\g$ theory should be precisely that of the Ising model.
Consider next a tree-level  $N$-point correlation function involving
$N_0$ vertices $V_0$, $N_\si$ vertices $V_\si$ and  $N_\ha$   vertices
$ V_\ha$  with $N=N_0+N_\si +N_\ha$. This will vanish unless   one
includes appropriate insertions  of the picture-changing operator $P$
(as in fermionic strings) and of the screening charges   (as in the
Coulomb gas construction). For any amplitude, it is sufficient to
include  $N_P$ picture-changing operator  insertions and $N_S$
insertions of the screening operator  $S$ to obtain  the free-field
correlation function $$
<\Omega|V_0(z_1)...V_0(z_{N_0})V_\si(w_1)...V_\si(w_{N_\si})
V_\ha(y_1)...V_\ha(y_{N_\ha})P^{N_P}S^{N_S}|\Omega>
\eqn\corr$$
where $|\Omega>$ is the product of the zero-momentum
 $\f$ Fock-space
vacuum and the   $SL(2,R)$
invariant vacuum of the ghost system.
The operators $V_0,V_\si,V_\ha,S,P$ have momenta
$8iq/7,iq,4iq/7,-2iq/7,0$ and
ghost-numbers $2,2,1,-1,1$ resepcectively.
 Momentum conservation
\foot{Strictly speaking, the
integration over the bosonic zero mode   does not give a
momentum-conserving delta function since the momenta are
imaginary. For the present purposes, we restrict
ourselves to the resonant amplitudes which conserve momentum.}
 in the presence
of a background charge $q$ implies that the amplitude vanishes unless
${1 \over 7} (8N_0+N_\si + 4N_\ha -2N_S)=2$, so that
the number of screening charges is fixed to be
$$N_S={1 \over 2} (8N_0+N_\si + 4N_\ha)-7 \eqn\nsis$$ For any
operator $X$, the
expectation value  $<0|X|0>$ will vanish unless the ghost number of $X$
is $5$; this can be thought of as the condition  that the insertion of
$X$ cancels the effect due to the anomaly in the conservation of
ghost-number. This will be the case if  $ 2N_0+2N_\si +  N_\ha -
N_S+N_P =5$, which together with \nsis\  fixes the number of
picture-changing operators to be  $$N_P=2N_0+{3 \over 2}N_\si +
3N_\ha -2 \eqn\npis$$ The amplitude is independent of the positions at
which the picture-changing operators are inserted, as  $$P(z_2)-P(z_1)=
[Q,K], \qquad K= \int _{z_1}^{z_2} dw  \pf (w) \eqn\ptriv$$ and the
insertion of a BRST-exact operator into a correlation function of
physical operators gives a vanishing result. Thus all that remains to
determine  the amplitude \corr\ is the choice of contours for the
screening charges $S$. Following [\dotso,\west], this is done by acting
with as many as possible of the factors of $S$ to convert the operators
\vis\ to ones of the form \vnis, and then using crossing symmetry and
unitarity to fix the remaining ones.

The calculations of all three-point and four-point correlation
functions are
contained as special  cases of the calculations  given in
[\pff,\pjj,\pooop,\west] and in all cases they agree with the
 corresponding Ising model  correlation functions.
 The Ising correlation function for $N_0$ identity operators,
$N_\si$ spin operators and $N_\ha$ energy operators is given by \corr\
where the number of screening charges and picture changing operator
insertions is fixed by \nsis,\npis. No other kinds of screening
operators appear to be necessary. In particular, the three-point
functions give   the correct fusion rules for the Ising model [\pff]
$$ \eqalign{ {\bf 1}\times {\bf 1}&={\bf 1}\ ,\qquad
\sigma\times\sigma={\bf 1}+\varepsilon\ ,\cr {\bf 1}\times \sigma&=
\sigma\
,\qquad \sigma\times\varepsilon=\sigma\ ,\cr {\bf 1}\times
\varepsilon&=\varepsilon\ ,\qquad
 \varepsilon\times \varepsilon= {\bf 1}\ .\cr}\eqn\isfus $$

As an example, consider the correlation function of four
spin-half operators.
This requires the insertion of one screening operator $S$ and two
picture changing operators.
Using one $\bar V(\ha , 0)=SP(V_\ha)$ operator and one $PV_\ha$
operator, it is straightforward to bosonise the ghosts and calculate
the correlation function  as in [\pff] to give
 $$\eqalign{
&<\Omega|V_\ha (z_1)V_\ha(z_2)
(SPV_\ha)(z_3)(PV_\ha)(z_4)|\Omega>
\cr &
=2(z_{12}z_{13}z_{14}z_{23}z_{24}z_{34})^{-1/3}x^{-2/3}(1-x)^{-2/3}
(1-x+x^2)
\cr}
\eqn\werewrjc$$
where
$$ z_{ij}=z_i-z_j, \qquad x= { z_{12}z_{34}\over
z_{13}z_{24}}
\eqn\cross$$
and this agrees with the corresponding Ising model correlation
function.

\chapter{$W_3$-Strings}

To attempt to construct a critical string theory, one can now add to
the  $\f ,\b ,\g $ realisation of the  Ising model an extra conformal
matter sector (whose fields we denote generically by $X$) with stress
tensor $T_X$ and the usual ghosts $b,c$ with stress tensor $T_{b,c}$.
The standard  BRST charge
$$Q_0= \int dz \, c\left(
T_X+T_\f +T_{\b, \g} + \ha T_{b,c}\right) \eqn\qois$$
 will then be nilpotent
provided that the total matter central charge is $26$, \ie\ provided
that the central charge $c_X$ for the matter system is $c_X= 25 \ha$.
Moreover,
this BRST charge anti-commutes with $\QQ$ so that
one can define the total BRST charge
$$\hat Q=Q_0+\QQ
\eqn\qtot$$
and this is iteslf nilpotent. The theory given by the matter sector $X,
\f$, the ghost sector $b,c,\b , \g$ and the BRST charge \qtot\ is
precisely the $W_3 $ string theory of [\wstr] (after a canonical change
of variables given in [\pooop]), with physical states identified with
the cohomology classes of \qtot.  Moreover, the matter system $(X,\f)$
is a conformal field theory with $c=100$ and  is a realisation of the
$c=100$ $W_3$ algebra constructed by Romans [\romans]; this is  the
most general $c=100$ realisation that  has so far been constructed and
so all known critical $W_3$ string theories  arise in this way. One
convenient choice of matter system is to take the fields $X$ to be $D$
free bosons with a background charge $a$, chosen so that $D+12 a^2 = 25
\ha$.

 If $V_\dd$ is any of the physical vertex operators of the
Ising system with dimension $\dd=0, \si$ or $\ha$
and $V_h$ is a primary field of the $X,b,c$ system with conformal
dimension $h$,
then the operator
$${\cal V}= cV_h(X,b,c)   V_\dd(\f,\b,\g)
\eqn\vtot$$ will   be $\hat Q$-BRST invariant provided $h+\dd=1$, so
that with $\dd=0, \si,\ha$ one can construct physical vertex operators
using vertices of the    $c=25 \ha$ matter theory corresponding to
states with intercepts
 $1,{15\over 16}$ and $\ha$.  There are also copies with $c$ replaced
with $c \pa c$. In the case in which the $X$ are free bosons,  all the
 physical states of the $W_3$ string in which the $X$
fields can have continuous momenta are obtained in this way, and this
has been
verified explicitly for the first few levels [\pcc-\west].

However, this does not exhaust the physical states of the theory.
Tensoring together the $X$ system and this realisation of the Ising
model leads to extra physical states at discrete values of the momenta
which are cohomology classes of $Q$ but do not factor into products of
Ising model states and $X$-states [\pff,\pcc,\ewrwe]. These arise in
exactly the same way as the extra states that were found in the Ising
model and can be understood in terms of singular vectors for the
algebra generated by $W$ and the total stress-tensor $T$ (instead of
just $T_\f+T_{\b,\g}$).

The identity operator $1$ is clearly BRST invariant, as is
its conjugate $c\pa c \pa^2 c \g\pa \g \pa ^2\g \pa^3 \g \pa ^4 \g$.
The picture-changing operator given by \pis\ is
not invariant under the total BRST charge $\hat Q$ and must be modified
to become $P'=[\hat Q, \phi] =P+c\pf -q \pa c$. It follows from \vnles\
that the operator $V_0=V(0,0)$ can be constructed from the identity
operator by $V(0,0)(z)= RP(1)=\oint dw \g(w) e^{-8q\f(w)/7}P(z) $  and
$cV_1(X)V_0$ will be BRST invariant for any $X$-space vertex operator
$V_1$ with weight $h=1$.
(Note that $cRP(1)=cRP'(1)$ as the extra term involves $cc=0$.)
However, one would expect to be able to construct
physical operators using the identity operator from the $X,b,c$
system and a dimension zero operator from the Ising sector.
The operator $V(0,0)= RP(1)$ is not invariant under the total BRST
charge, but is invariant if $P$ is replaced by $P'$.
This gives a new physical operator $RP'(1)=
(c \g+{2 \over 87}\sqrt{58}i\pa \g \g)e^{-8q/7}$, which is one of the
discrete state vertex
operators found in [\ewrwe].
An infinite sequence of physical discrete states
involving the identity operator from the
$X,b,c$ sector and a dimension zero operator from the Ising sector can
be constructed from the identity operator by acting with
$S,P',R,\bar R$ on the identity operator:
$$\eqalign{&
V'(0,n)=[S^3 P'SP']^{n-1}(1), \qquad \bar V'(0,n)=SP'V(0,n)
\cr &
V'(0,-n)=[RP']^{n+1}
(1)
\cr}
\eqn\vpisis$$
These reduce to $V(0,n),\bar V(0,n)$ on replacing $P'$ with $P$, and
are
manifestly invariant under the total BRST charge $\hat Q$, since
$S,R,P'$  are.
The identity operator  is a
discrete state of the $X,b,c$ system and if the
$X,b,c$ system has extra discrete states, as will be the case
if the $X$ system consists of a single boson with background charge,
then there will be extra discrete states constructed by replacing
$1$ in \vpisis\ with the operators corresponding to these discrete
states. All of the discrete states  so far constructed for the
$W_3$ string arise in this way.

Of particular interest is the case in which the extra matter fields
consist of a single extra boson $X$ with background charge. The theory
so constructed is now seen to be precisely the Ising model  coupled to
two-dimensional gravity (\ie\ Liouville theory) and this is known to
have an infinite  number of extra discrete physical states.
This model can also be viewed as the critical  $W_3$ string
constructed from the two-boson $c=100$ realisation of $W_3$; this model
and its discrete state structure has been analysed from the
$W_3$ point of view in
 [\ewrwe].
{}From the present point of view,
these arise because the $X,b,c$ system has an infinite number of
discrete states in its
own right, and new discrete states
are obtained by tensoring together discrete states from the $X,b,c$
sector with ones from the $\f ,\b,\g$ sector, after replacing $P$ with
$P'$.

For $D$ bosons $X$, the model becomes equivalent to a system
of $D-1$ free bosons without background charge (so their contribution
to the central charge is $D-1$) plus the $c=\ha$ Ising model coupled
to two-dimensional gravity, so that it could well be the case that
some of the work on $W_3$ strings could shed light on the coupling
of two-dimensional gravity to $c>1$ matter; in particular, the
$W_3$ string analysis shows that
this theory must also have extra discrete states,
but not as many as for the $c \le 1$ matter theories.

\chapter{Other W-Strings and Other Minimal Models}

The discussion of the relation between the $W_3$ string and the Ising
model
  generalises to other $W$-string   theories [\pbb,\lat] and minimal
models. A simple generalisation of \lis\ is the lagrangian
$$L= \pf \bar \pa \f + B\tilde W, \qquad \tilde W= {1 \over s}( \pf)^s
\eqn\lisn$$
for a free boson subject to the   constraint $( \pf)^s \sim 0$,
with $B$ a spin-$s$ gauge field. Quantising the theory in the gauge
$B=0$ requires the introduction of an anti-commuting ghost $\g$ of
spin $1-s$ and an anti-ghost $\b$ of spin $s$.
The naive BRST charge is of the form
$Q=\int dz \g \tilde W+...$ and as in the $s=3$ case it must be
modified by higher derivative terms for $Q$ to be nilpotent.
The results of [\lat] give an explicit construction of a
nilpotent BRST charge for $s=4,5,6$ and there is an implicit
construction of $Q$ from the BRST charge for $W_s$ strings for all $s$.
The stress-tensor is
 of the form \ttis\
where
$$T_{\f}=- {1\over 2} \pa \f \pa \f - q \pa ^2 \f,
\qquad
T_{\b \g} = -s \b \pa \g -(s-1) \pa \b \g
\eqn\tisis$$
and $q$ is a background charge, which is determined by requiring
$[Q,T]=0$ and given by [\lat]
$$q^2={(s-1)(2s+1)^2\over 4(s+1)}
\eqn\qiss$$
As in the spin-3 case, there is a $W_\infty$ symmetry generated by
currents
$W^{(n)}=(\pf)^n+...$, $n=2,3,4,...$ which commute  with the
BRST operator. In this case, the  currents of spin $n \ge s$
are BRST trivial, and so topological, while the
currents of spins $2,3,...,s-1$ are BRST non-trivial.
In particular,
$W^{(s)}=\{Q,\b \}$.

 The  total central charge of the $\f,\b,\g$
system is
$$
c_s=1+12 q^2-2(6s^2-6s+1)={2(s-2)\over s+1}
\eqn\cwis$$
The $N$'th minimal model $M_{M,N}$ of the $W_M$
algebra has central charge
$$c_{M,N} =(M-1)\left(1- {M(M+1) \over N(N+1)}
\right)\eqn\cnmis$$
and so \cwis\ is $c_{s-1 ,s}$, suggesting that this model
could be the first minimal model of the $W_{s-1}$ algebra.
Defining the physical states to be the cohomology classes
of the BRST operator $Q$ gives a model with a discrete spectrum
which
  the analysis of [\lat] shows
to be consistent with the spectrum of the first $W_{s-1}$
minimal model. Much of the analysis of this paper carries over
straightforwardly to this case.

The quantisation of the system \lisn\ outlined above is not
 unique; for $s=4$ there are three   nilpotent BRST charges with the
correct classical limit,   for $s=6$ there are four, while for $s=3$
or $s=5$ there is only one [\lat]. The different BRST charges
correspond
to different values of the background charge and hence of the central
charge $c$. In addition to the models described above which exist for
any $s$, there are then certain extra models which are not yet
understood, but which may correspond to non-unitary minimal models.

As in the case $s=3$, a critical string theory can be constructed by
adding an effective matter system $X$ and ghosts $b,c$
to construct a total BRST charge
\qois,\qtot\ which will be nilpotent provided the central charge of
the effective matter system is $26-c_s$. This gives the W-string
theories of [\lat], based on the quantization of certain $W$-gravity
theories first constructed in [\hee],
which are seen to consist of $c=26$ strings which
include a $W_{s-1}$ minimal model in its conformal matter sector.

Another interesting class of models is found by considering   the
construction of $W_N$ strings [\pbb].
The $W_N$ algebra is  generated by the spin-two
stress tensor $T$ and currents $W^{(s)}$ of spins $s$, $s=3,...,N$. The
construction of critical $W_N$ string theories requires a matter sector
which is a realisation of the $W_N$ algebra whose central charge takes
the critical value
 $$c_N^*=2(N-1)(2N^2+2N+1)\eqn\cnn$$
together with
the usual conformal ghosts $b,c$ plus the W-ghosts $\b^{(s)} ,\g^{(s)}$
of spins $s,1-s$, $s=3,...,N$.
One
$W_N$  matter  realisation with critical central charge \cnn\ [\pbb]
is given by taking an arbitrary conformal
 field theory $X$ with central
charge
$$\tilde c= 26 - \left(1 - {6  \over N(N+1)} \right)= 26- c_{N-min}
\eqn\ceffna$$
and adding $N-2$
free bosons $ \phi _s$ ($s=3, \dots ,N$)
with background charge. The system given by these $N-2$ scalars
and the ghosts $\b^{(s)}, \g^{(s)}$ for $s=3,4,...,N$ has central
charge  $c_{N-min}$.
A BRST charge has not been constructed explicitly for $N>3$, but it
seems reasonable to expect that such charges exist,
and that they can be written in   the form $\hat Q=Q_0+Q_N$, where
$Q_0$ is the usual Virasoro BRST operator \qois\ and $Q_N$ acts
only on the $(\f_s,\b_s,\g_s)$ system.
Then the model defined by restricting the $(\f_s,\b_s,\g_s)$ system
to the cohomology classes of $Q_N$ is
a conformal field theory with the central charge of the $N$'th
Virasoro minimal model and
the results
of [\pbb] for the spectrum of this model
are consistent with the identification of this model with the $N$'th
Virasoro minimal model.
Then the $W_N$ string based on this realisation of $W_N$
is, at tree level, a $c=26$ string whose matter sector consists
of the $N$'th Virasoro minimal model, plus an arbitrary
conformal field theory with central charge \ceffna.

There are undoubtedly many similar free boson and ghost models
which correspond to minimal models. As far as I am aware,
all known free-field representations of $W$-algebras are of the
form $(X,\f)$, where $X$ are the fields of some conformal field theory
on which the higher-spin W-generators do not act,\foot{\ie\ there is a
non-primary basis for the generators of the W-algebra for which this is
so.}
 plus some extra bosons $\f$, which have the
property that combining the bosons $\f$ with the $W$-ghosts $\b , \g$
(and truncating using a BRST operator) gives a minimal model. Using
such a W-representation to construct  a W-string theory  will always
give a normal Virasoro string which includes a minimal model sector,
with the W-symmetry acting only within the minimal model sector. It
would be interesting to find representations of W-algebras which are
not reducible in this way, as they would offer the prospect of truly
new W-strings.

\

\ack

I would like to thank J. Distler, B. Spence and C. Thorn for helpful
discussions and the ITP for its hospitality.
This work was  supported by NSF grant no. PHY 89-04035.

%\vfill
%\eject

 \refout
\end